\documentclass{osa-article}

\journal{osac}

\articletype{Research Article}

\begin{document}

\title{Detection of Proteoglycan Loss from Articular Cartilage using Brillouin Microscopy, with Applications to Osteoarthritis}

\author{Pei-Jung Wu\authormark{1,3}, Maryam Imani Masouleh\authormark{2}, Daniele Dini\authormark{2}, Carl Paterson \authormark{3}, Peter T{\"o}r{\"o}k\authormark{3,4}, Darryl R. Overby\authormark{1*},  and Irina V. Kabakova\authormark{3, 5**}}

\address{\authormark{1} Department of Bioengineering, Imperial College London, London, UK\\
\authormark{2} Department of Mechanical Engineering, Imperial College London, London, UK \\
\authormark{3} Department of Physics, Imperial College London, London, UK \\
\authormark{4} Division of Physics $\&$ Applied Physics, Nanyang Technological University, Singapore, Singapore \\

\authormark{5} School of Mathematical and Physical Sciences, University of Technology Sydney, Sydney, NSW, Australia}

\email{\authormark{*}d.overby@imperial.ac.uk} 
\email{\authormark{**}irina.kabakova@uts.edu.au} 


\begin{abstract}
The degeneration of articular cartilage (AC) occurs in osteoarthritis (OA), which is a leading cause of pain and disability in middle-aged and older people. The early disease-related changes in cartilage extra-cellular matrix (ECM) start with depletion of proteoglycan (PG), leading to an increase in tissue hydration and permeability. These early compositional changes are small (<10$\%$) and hence difficult to register with conventional non-invasive imaging technologies (magnetic resonance and ultrasound imaging). Here we apply Brillouin microscopy for detecting changes in the mechanical properties and composition of porcine AC. OA-like degradation is mimicked by enzymatic tissue digestion, and we compare Brillouin microscopy measurements against histological staining of PG depletion over varying digestion times and enzyme concentrations. The non-destructive nature of Brillouin imaging technology opens new avenues for creating minimally invasive arthroscopic devices for OA diagnostics and therapeutic monitoring.   
\end{abstract}

\section{Introduction}
AC is a highly organized connective tissue, comprising a single type of specialized cell - the chondrocyte - within an ECM \cite{Mansour}. The structure and arrangement of cartilage components are organized to serve the tissue's main function of load bearing, resilience to mechanical wear and redistribution of stresses in order to protect the underlying bone. A variety of complex interactions between ECM and the chondrocytes maintain a fine balance between AC synthesis and degradation \cite{Goldring}. Matrix-degrading molecules, including matrix metalloproteinases (MMPs) and enzymes such as aggrecanase, are produced by chondrocytes under normal and pathological conditions. Abnormal load distribution, accelerated mechanical wear of the cartilage surface and overproduction of matrix-degrading components all can trigger pathological processes in chondrocytes and ECM, leading to disease and irreversible degradation of articular cartilage \cite{Mansour,Goldring, Hendren}.

The main components of articular cartilage are water (70$\%$ to 85$\%$ of weight) and the ECM, which is composed of type II collagen (15$\%$-20$\%$ of weight) and proteoglycans (PGs) (3$\%$-10$\%$ of weight)\cite{Matzat}. The protein cores of PGs are lined by covalent attachments of glycosaminoglycans (GAGs), which confer negative charge due to the abundance of carboxyl and sulfate groups.  This property fixes PGs to the ECM and attracts cations, such as sodium, which then draw water into the tissue to generate the swelling pressure of cartilage. The ECM network generates a resistance to interstitial fluid flow, which determines the rate of tissue deformation. Thus the structure of ECM provides AC its load bearing function \cite{Borges, Accardi}. 

OA is a painful joint disease associated with breakdown of articular cartilage and underlying bone \cite{Goldring}. The progression of OA is classified into three stages: 1) proteoglycan degradation followed by degradation of type II collagen, 2) the fibrillation and erosion of the cartilage surface, and 3) the onset of the synovial inflammation \cite{Matzat}. This process is schematically illustrated in Figure 1 for healthy cartilage (a), early (b) and late (c) stages of OA. The progression starts from the molecular level (PGs depletion), evolves to the architectural changes within ECM (collagen network erosion) and ends up at irreversible structural and functional damage. 

The PG depletion leads to  increased tissue permeability and decreased fixed charge density that reduces the load-bearing function of AC. The increased permeability, in turn, reduces the capability of the fluid to support load and causes higher stresses in the ECM, hence triggering more mechanical damage \cite{Borges}. Because the progression of AC degradation often occurs over several years, detection of OA during the early stages of the disease is crucial and limits disease severity \cite{Matzat}. 

Conventional approaches to OA diagnostics include qualitative grading techniques based on arthroscopic and x-ray images \cite{Wright}. These techniques are sensitive to late stages of OA when cartilage thinning and visible lesions are apparent (Fig. 1(c)). Quantitative high-resolution magnetic resonance imaging (MRI) \cite{Matzat, Braun} and atomic force microscopy (AFM) nano-indentation \cite{Stolz} have demonstrated ability to detect early stages of OA, however simpler, cheaper and more reliable solutions are still required, in particular for {\it in vivo} and {\it in situ} diagnostics.  

\begin{figure}[ht!]
\centering\includegraphics[width=12cm]{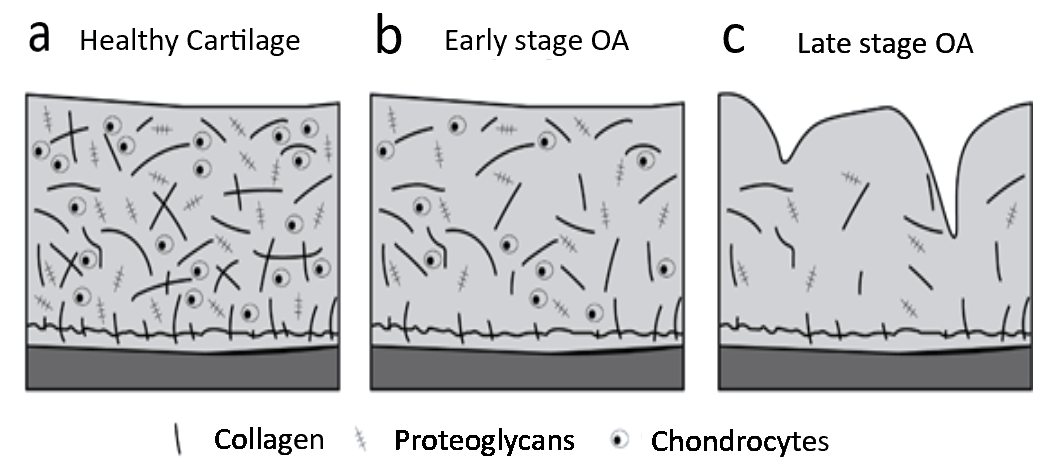}
\caption{Representation of articular cartilage composition at three stages of OA progression: (A) healthy cartilage, (B) early stage OA (C) late stage OA (image is adopted from \cite{Matzat}).}
\end{figure}
%
%
Brillouin microscopy (BM), a non-destructive and label-free technique \cite{Koski}, has been suggested for biomechanical assessment of biomaterials \cite{ScarcelliNature}, including tendons, ligaments and cartilage \cite{Palombo}. BM is based on optical detection of acoustic waves inside the material and can relate the speed of sound of these waves to mechanical parameters such as longitudinal modulus, compressibility and viscosity \cite{Cusack,Harley}. The spatial resolution of this technique is similar to that of confocal microscopy and can reach sub-micron scale, enabling measurements of cellular and sub-cellular structures \cite{Scarcelli_NatMeth, sub-cellular}. We have recently shown that BM measurements of hydrated materials are highly sensitive  to hydration level \cite{Wu}. Since the change of cartilage hydration level is the consequence of PG depletion, we hypothesize that BM can be used to monitor the PG depletion at the early stage of OA disease. 

Here we apply BM to explore the relationship between PG depletion and changes in Brillouin frequency shift in porcine articular cartilage exposed to enzymatic treatment. We compare BM results against histological staining of PG and observe qualitative agreement between histology and Brillouin microscopy measurements. Future developments in flexible Brillouin fiber probes \cite{Kabakova_endoscope} fitted within traditional or novel arthroscopy devices can lead to minimally invasive diagnostics and monitoring of early-stage OA. 

\section{Methods}
\subsection{Sample preparation}

10~mm diameter biopsy punches containing articular cartilage and underlying bone were obtained from 2 porcine shoulder joints (5 punches per joint, including one for control sample and 4 for digestion experiment) sourced from a local abattoir within 24 hours of slaughter. Punches were incubated in 0.1 or 1 mg/ml trypsin at 37$^o$C for 1-4 hours. After enzymatic digestion all 10 samples were washed in phosphate buffered saline (PBS) to quench the digestion process and sectioned in 2 equal parts using a surgical scalpel along the plane perpendicular to the articular surface. One half of the punch was used for Brillouin microscopy measurements and the other half for histology. All samples were soaked in PBS for storage and kept refrigerated until the measurements.
%
%

\subsection{Histology}
For histology measurements the tissues were fixed in 10$\%$ natural buffer formalin (NBF) for 48 hours straight after digestion. Following fixation, the samples were decalcified with 10$\%$ formic acid, dehydrated through a graded ethanol concentration (50$\%$, 70$\%$, 90$\%$ and 100$\%$), cleared using histoclear for 10 min, infiltrated and embedded in paraffin wax. 0.001$\%$ Fast Green stain was applied for 6 min to color the bone and 0.1$\%$ Safranin O was used  to stain PGs (2 min). Five-micrometer-thick sections were cut using the microtome (Leica).

\subsection{Brillouin microscopy}
Brillouin microscopy was performed on fresh samples within 24 hours after enzymatic treatment. Samples were placed in a glass-bottom dish with the articular surface facing down and immersed in PBS. A spacer was inserted between the sample and the bottom of the dish to prevent direct contact of the sample and the dish. The construction of the Brillouin microscope was analogous to previously reported \cite{Wu} and consisted of a single longitudinal mode laser (Cobolt, 561~nm, 30 mW), a confocal microscope, an interferometric filter to reduce unwanted peaks at the laser frequency and a 1-stage virtually-imaged phase array (VIPA) spectrometer \cite{AntonacciArtery, AntonacciFilter}. An objective lens with numerical aperture NA=0.5 was used to obtain a lateral and axial spatial resolution of approximately 1 and 3~$\mu$m, respectively. The choice of moderate numerical aperture is to avoid possible spectral broadening of Brillouin peaks characteristic of imaging with high numerical aperture objectives (NA>0.7) \cite{Antonaccibroadening}.

The sample was positioned on an X-Y-Z motorized stage. By moving the sample in the vertical direction, we obtained a linear Z-scan of Brillouin frequency shift in the direction perpendicular to the cartilage articular surface. Each Z-scan started approximately 100~$\mu$m lower than the cartilage articular surface and went upwards with steps of 2, 4, or 10~$\mu$m each, and five spectra were taken for each step. We also performed Brillouin microscopy measurements in planes parallel to Z axis (YZ-scans). The step size for movements along Y and Z axis  was chosen to be 10~$\mu$m/step.  
%
%

Raw Brillouin spectra were processed using a custom-built MATLAB (MathWorks) code. Briefly, two brightest diffraction orders were chosen from each image collected by the camera. The images were then converted to the frequency coordinate by least-square fitting using a quadratic function \cite{Wu}. The Brillouin frequency shift was obtained by averaging five spectra taken from the same spatial location and fitted with a Lorentzian function. The latter took into account the linear background from the Rayleigh scattering.


The output of the Brillouin microscopy measurement gives the frequency shift between the Rayleigh light (elastically scattered) and Brillouin light (inelastically scattered). This frequency shift, called Brillouin frequency shift $\Omega$, is equal to the frequency of the longitudinal acoustic mode inside the material, and in our experimental geometry it is proportional to the square root of the longitudinal elastic modulus 
\begin{equation}
\Omega=\frac{2n\pi}{\lambda}\sqrt{\frac{M}{\rho}},
\end{equation}
where $n$ is the refractive index of the medium, $\lambda$ is the wavelength of the laser, $M$ is the material's longitudinal modulus and $\rho$ its density \cite{Lepert}.

\subsection{Statistical analysis}
A two-way mixed ANOVA was used to test the effects of trypsin concentration and digestion time on Brillouin measurements (SPSS Advanced Statistics module, version 20; IBM). The within-subject factor was defined to be Brillouin relative frequency shift, whereas between-subject factors were defined as the trypsin concentration (0.1 mg/ml or 1 mg/ml) and digestion time. After the two-way mixed ANOVA, we used a post-hoc univariate test to detect significantly different Brillouin measurements between concentrations at each time point.

\section{Results}
\subsection{Control measurements prior enzymatic digestion}

The biophysical structure of cartilage varies spatially in lateral and axial directions \cite{Matzat}. To study natural variation in $\Omega$ related to the cartilage anisotropy, we first measured $\Omega$ along the articular surface of the untreated, "control"  AC sample. Figure 2 presents the results of this test for (a) varying step size of Z-scans and (b) varying position along the articular surface. 

The results in 2(a) show no difference in $\Omega$ depending on the step size, as expected. The penetration depth for these measurements was found to be approximately 100 $\mu$m. The origin of the Z-axis for these measurements has been chosen arbitrarily inside the PBS solution, but was the same for each scan and sufficiently close to the articular surface.   
%
%

The fairly small penetration depth can be explained by the loss of light due to multiple scattering events in the turbid cartilage tissue. Improving microscope sensitivity, for example by adding extra filters that reduce Rayleigh light or by switching to a different type of spectrometer (e.g. a tandem Fabry-Perot scanning interferometer \cite{Sandercock}) will enable detection of Brillouin signals from deeper layers. Currently, however, the shallow penetration depth limits analysis of the outermost layers of the cartilage structure, including the superficial and tangential zones. 

%
%
\begin{figure}[h!]
\centering\includegraphics[width=14cm]{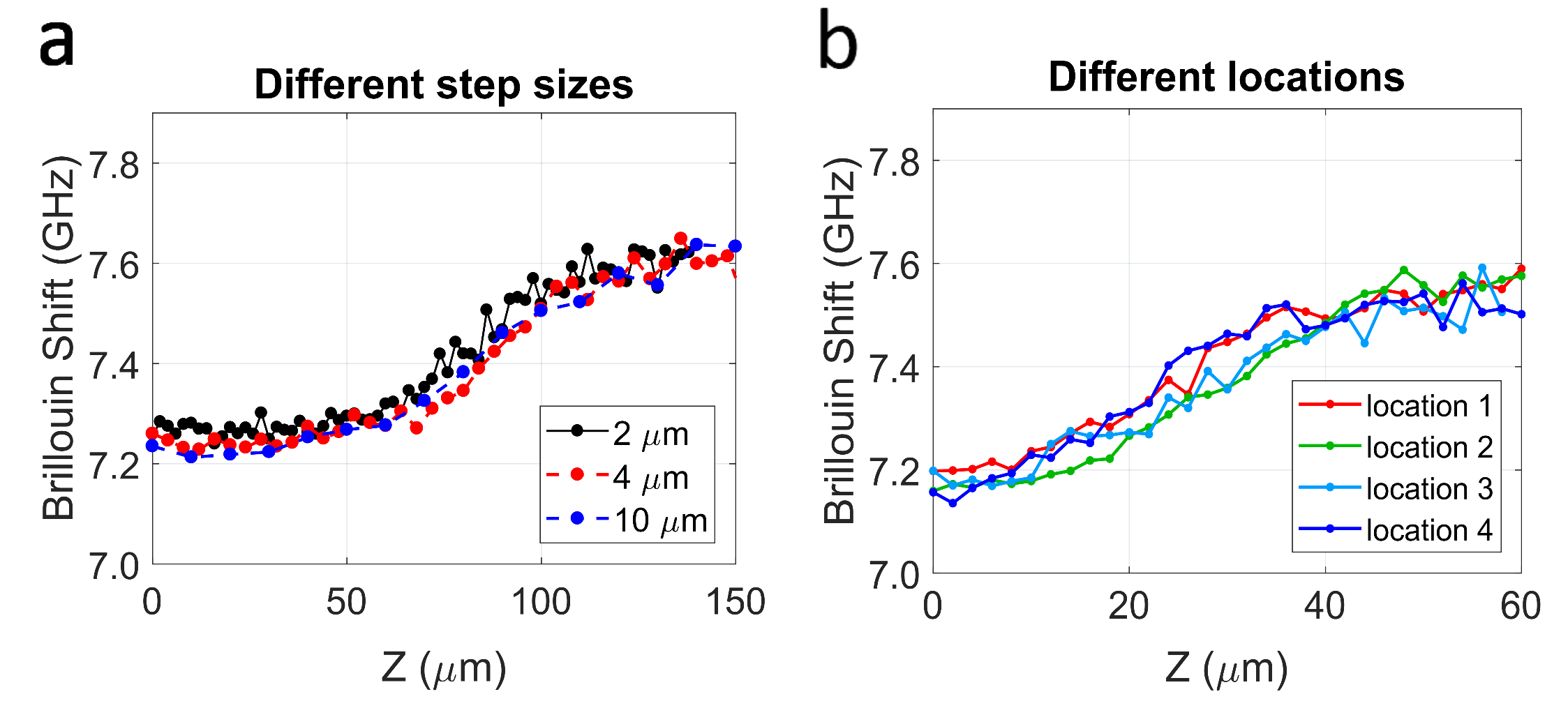}
\caption{ Brillouin microscopy scans in the direction perpendicular to the articular surface of cartilage (Z-scans) for: a) different step sizes of 2, 4 and 10 $\mu$m and b) different locations along the articular surface. }
\end{figure}

Changing the imaging location resulted in no significant change in the trend of $\Omega$ across the articular surface, with the standard deviation at each scan depth Z being well within  0.06~GHz  (Fig. 2 (b)). Locations 1-4 in Fig. 2(b) were approximately 2 mm apart from each other. These data indicate relatively little variability in the Z-dependence of $\Omega$ across the cartilage surface. 

All measurements presented in Figure 2 (a-b) demonstrate a smooth transition in $\Omega$ from 7.2~$\pm 0.03$ GHz (in PBS solution) to approximately 7.6~$\pm 0.03$ GHz (in cartilage). The existence of a smooth transition rather than a step transition may be attributed to a gradual change in composition of the cartilage ECM. 

\subsection{Trypsin treatment}
\begin{figure}[h!]
\centering\includegraphics[width=14cm]{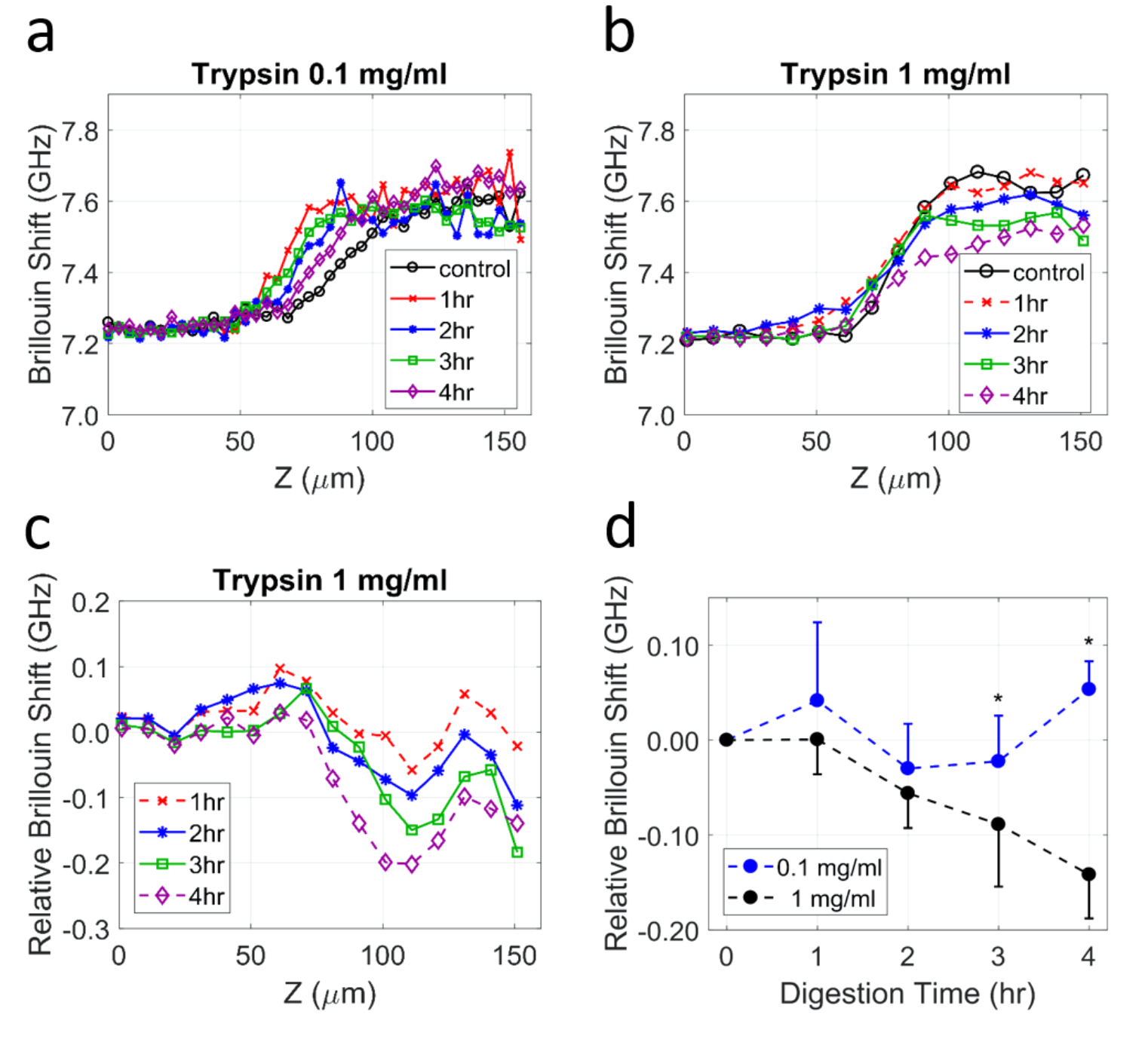}
\caption{Brillouin microscopy of digested cartilage samples using a) 0.1 mg/ml and b) 1 mg/ml trypsin solution. Z-scan of the relative Brillouin frequency shift compared to the control sample is demonstrated in c) and the average relative shift measured inside the cartilage at Z>150~$\mu$m in d). Asterisks signify time points with statistically significant variation in the relative frequency shift (see Section 2.4). }
\end{figure}

Articular cartilage samples were treated with 0.1 and 1 mg/ml trypsin solution and $\Omega$ was measured as a function of depth  after 0 (control), 1, 2, 3 and 4 hours of enzymatic digestion. Results of BM measurement for the two treatment scenarios are shown in Figure 3 for the absolute value of $\Omega$ (a-b) and the relative change in $\Omega$ compared to the control sample (c-d). 

For samples treated with 0.1 mg/ml trypsin solution (Fig. 3(a)), the maximum variation in $\Omega$ inside the cartilage (Z>100~$\mu$m) is within 0.09~GHz, slightly higher than the measurement uncertainty for control samples (0.06~GHz). There is, however, no clear dependency on treatment time for 0.1 mg/ml digestion as evident from Figure 3(d). The relative Brillouin frequency shift fluctuates about 0~GHz, suggesting that there are no appreciable changes in $\Omega$ compared to the control sample during the 4 hour digestion period. A higher variability in $\Omega$ for locations inside the digested cartilage compared to those in the control sample   can be explained by stronger light scattering in the digested tissue. We hypothesize that micro-scale damage induced by enzymes leads to  irregularities in the cartilage structure and hence stronger elastic scattering of light inside the tissue can be expected. The increase in light scattering, in turn, results in higher signal attenuation, weaker collected signal and a decrease in the signal-to-noise ratio, explaining the larger uncertainty in the measurements of the digested versus control tissue. 

For samples treated with 1 mg/ml trypsin, we observed a clear decrease in the maximum $\Omega$ inside the cartilage tissue with increasing digestion time (Fig. 3(b) and 3(d)). With increasing time of digestion, the relative Brillouin frequency shift between the trypsin treated cartilage and the control sample grows to a maximum of 150~$\pm$ 0.045~MHz as detected after t=4 hours of treatment (Figure 3(d)). This change exceeds the measurement uncertainty and hence is likely related to the structural and compositional changes taking place during the digestion process.

We applied a two-way mixed ANOVA to further examine the relationship between enzyme concentration and digestion time (Section 2.4). There was a statistically significant interaction between the two concentrations and time (p<0.0005). The post-hoc test confirmed a statistically significant difference in the relative Brillouin shift at the third (p=0.013) and the fourth hour (p<0.0005) with 1 mg/ml trypsin solution. 

\subsection{Histology results}
To confirm that the change in $\Omega$ is associated with a loss of solid ECM components, we performed histological assessment of PGs and GAGs on porcine AC following exposure to trypsin (see Section 2.2 for details). Histology was performed using Safranin O that stains polyanionic GAGs. The histology results (Figure 4) show a significant reduction in Safranin O staining already within 1 hour of digestion treatment. The reduction in PG labeling was most apparent at the articular surface. With increasing digestion time, PG depletion extended towards the cartilage-bone interface. Since the stain targets GAG molecules in PG component of the cartilage ECM, we can connect the decline in stain concentration with progressive depletion of GAG molecules as the result of trypsin digestion. The GAG depletion would also lead to higher tissue permeability and fluid influx into the tissue. This could account for the reduction in $\Omega$ that we observed for enzyme-digested samples, consistent with our hypothesis. 

\begin{figure}[h!]
\centering\includegraphics[width=13cm]{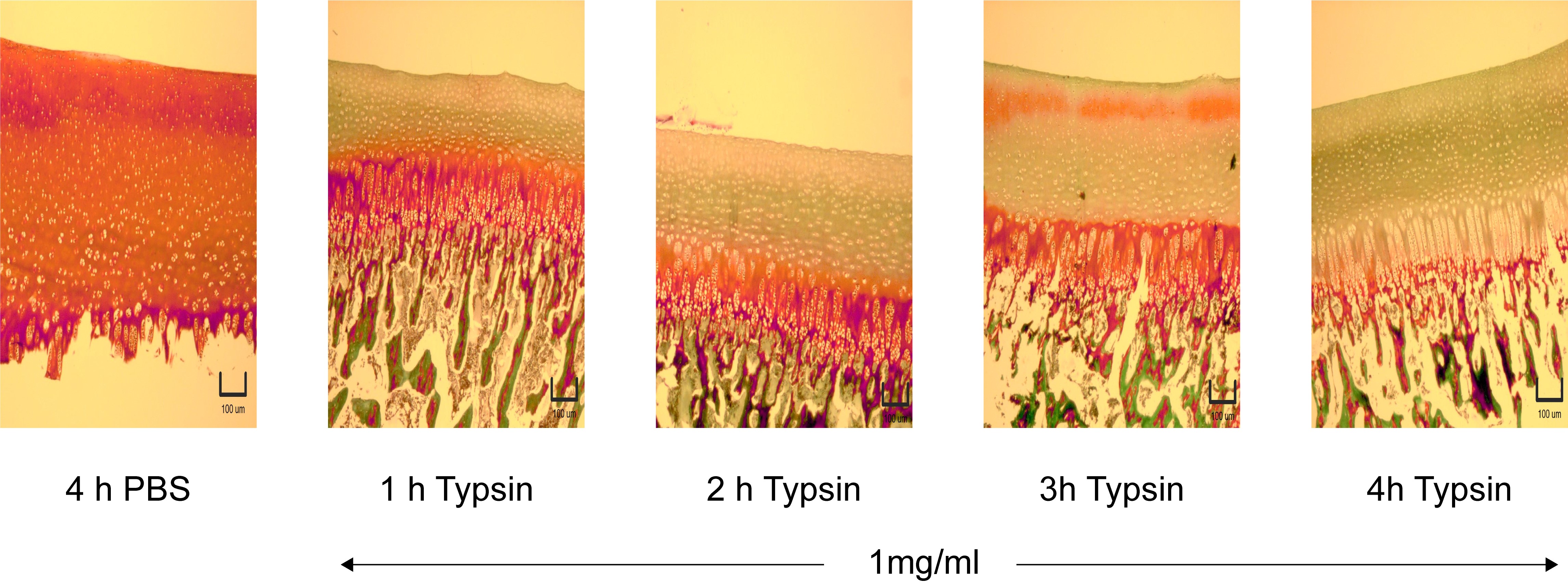}
\caption{Safranin O staining of articular cartilage reveals progressive GAG depletion after 1, 2, 3 and 4 hours of trypsin (1 mg/ml) digestion versus control sample (left).}
\end{figure}

\subsection{Trypsin diffusion uniformity assessment}

To assess the uniformity of trypsin diffusion into the cartilage, we partially digested cartilage punches attached to the bone in a 1 mg/ml trypsin solution for 1 hour. Then we separated the cartilage tissue from the bone and cut a tissue section sized around 10mm x 2mm x 2mm in the direction perpendicular to the articular surface. The 2D map of $\Omega$ collected from 140$\mu$m x 28$\mu$m area of this section is presented in Figure 5(a). Sectioning the tissue along Z-axis enables us to map Brillouin signal from deeper cartilage layers that are normally unreachable due to shallow penetration depth of the measurement technique.    

During imaging the section was immersed in PBS to avoid tissue drying, hence $\Omega$=7.2~GHz at Z=0~$\mu$m and Z>120~$\mu$m corresponds to PBS solution. The direction of Z axis in this graph is from the articular surface towards the bone. Following this direction, we notice that $\Omega$ first rises to about 7.6 GHz, then plateaus at this value and finally grows to 7.8 GHz at z=100~$\mu$m, before dropping back to the initial value of 7.2 GHz (PBS). This dynamics is consistent for any chosen position along Y axis, suggesting that the digestion process was relatively uniform across the entire sample and that the superficial layers of cartilage exhibit stronger GAGs depletion than the deeper layers. This is, perhaps, not surprising if to take into account that the digestion treatment was applied to the whole sample with the attached bone. Thus the diffusion of trypsin was directed from the articular surface rather than from the subchondral bone (Fig. 1).

\begin{figure}[h!]
\centering\includegraphics[width=14cm]{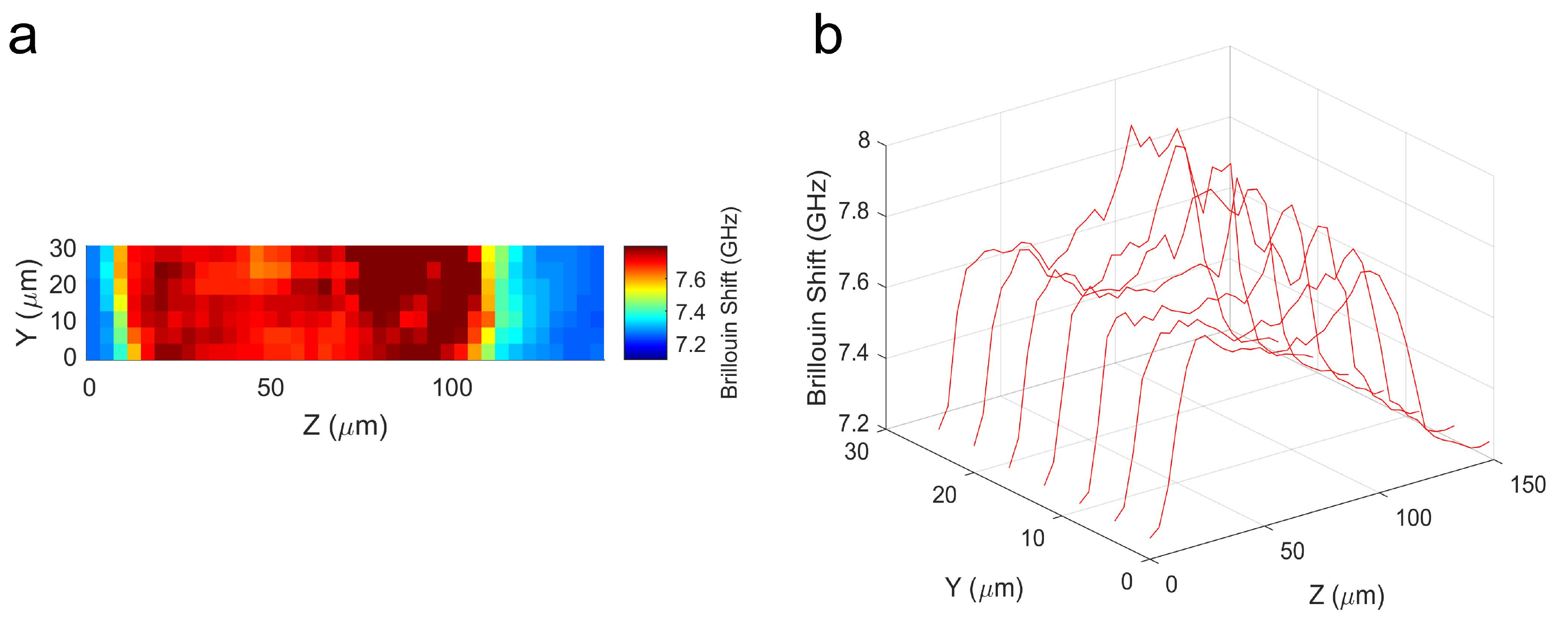}
\caption{Brillouin map of an articular cartilage section (140$\mu$m x 28$\mu$m), digested in a 1 mg/ml trypsin solution for 1 hour: a) full data and b) line representation with scans taken at 4 $\mu$m steps in Y-direction. }
\end{figure}

\section{Discussion}
  
Articular cartilage is no longer considered to be an inert tissue that is inevitably damaged as a result of wear and tear or of aging. Instead it is proving to be a unique, dynamic and specialized tissue capable to restoring and remodeling itself \cite{Hendren}. The conditions in which articular cartilage can undergo self-restoration are largely unknown, although clinical evidence suggests that moderate exercise could play a positive effect on stimulation of chondrocytes to produce new ECM components (PGs and collagen fibers) \cite{Mansour}. 

New assessment methods capable of detecting small variations in the biochemical and structural composition of cartilage ECM can be extremely valuable in finding and refining the recipe for maintaining healthy articular cartilage. One of the markers that signals about potential problem in cartilage health is increase in the ECM water content \cite{Buckwalter}. It has been shown using quantitative MRI imaging, specifically T2 mapping,  that increase of ECM permeability to water causes an elevation in both T2 relaxation time and the amount of free water in AC \cite{Matzat}.          

Previously we have found that BM is highly sensitive to the biomaterial's water content and have suggested that BM technique has potential in assessment of the local hydration level in tissues and cells \cite{Wu}. In a two-phase material consisting of a liquid ($\epsilon$) and a solid ($1-\epsilon$) components, with liquid component being dominant, the aggregate longitudinal modulus can be approximately described by a relationship
\begin{equation}
\frac{1}{M}=\frac{\epsilon}{M_l}+\frac{1-\epsilon}{M_s},
\end{equation}
where $M_{l,s}$ are the longitudinal moduli of liquid and solid parts, respectively \cite{Wu}. Since cartilage consists of mostly liquid phase (70-85~$\%$), Eq.~(2) can be applied in the analysis of BM results measured on control and digested cartilage samples. 

In the approximation of high liquid content and the solid component being less compressible than the liquid or $\frac{M_s}{M_l}>>1$, we arrive to an approximate relationship for the changes in the longitudinal modulus due to the depletion of solid component
\begin{equation}
\frac{\Omega^2}{\Omega^2_{0}}=\frac{M}{M_{0}}\approx\frac{\epsilon_{0}}{\epsilon}.
\end{equation}
Here $\Omega_{0}$, $M_{0}$ and $\epsilon_{0}$ correspond to Brillouin frequency shift, the longitudinal modulus and the hydration level of the control AC, whereas $\Omega$, $M$ and $\epsilon$  are parameters of the digested tissue. This approximation is valid for articular cartilage as dry ECM network was found to have significantly higher longitudinal modulus ($M_{s}=13.4$~GPa) than the bulk modulus of water ($M_{l}=2.2$~GPa) \cite{Palombo}. From Figures 3~(d) and 3~(e) we conclude that the maximum detected changes in $\Omega$ after 4 hours of digestion with 1 mg/ml trypsin is 150~MHz. According to Eq. (3) this suggests a relative increase in water content of approximately 4~$\%$, quite a realistic estimate of the liquid/solid fraction changes during the digestion process. It is worth noting, that current BM setup has the uncertainty of approximately 60 MHz, associated with the instrument noise level and the efficiency of Rayleigh peaks filtering, thus changes in water content of approximately 2~$\%$ can be detected using our system. Such a precision is almost an order of magnitude better than that achievable with both quantitative MRI \cite{Matzat} and AFM nano-indentation methods \cite{Stolz}. 

It is worth noting that the current model governed by Eq. (3) is a rather simplified picture of the digestion process. First, the changes in the cartilage content may lead to changes in the refractive index and the density of the cartilage, which also will affect the measurement of $\Omega$ according to Eq. (1). Secondly, the digestion products that remain in the tissue and may change the properties of PBS solution have been ignored at this stage, but their effect needs to be investigated in the future.  

\section{Conclusion}

In conclusion, we have applied Brillouin microscopy for studying the relationship between the tissue's longitudinal modulus and the depletion of a solid component of the ECM in porcine AC exposed to enzymatic digestion. Trypsin treatment of cartilage tissue leads to depletion of GAG molecules and increased liquid content mimicking  biochemical, compositional and structural changes in early stages of OA disease. We have observed qualitative agreement between BM measurements and histology analysis of samples digested with 1 mg/ml trypsin over 1-4 hours. The maximum change of 150 MHz in Brillouin frequency shift can be related to 4~$\%$ variation in the ratio between liquid and solid phases of the ECM. This suggests that BM is capable of detecting a few percent variation in the local hydration of the tissue, being at least an order of magnitude higher in sensitivity compared to quantitative MRI technique. Combining Brillouin imaging with modern arthroscopy can provide a technology resulting in a new minimally-invasive diagnostic tool for early-stage OA. Further studies, however, especially using human tissues and {\it in vivo} clinical studies are needed to validate this hypothesis.    

\section*{Funding}
I.~V.~Kabakova acknowledges financial support received through Junior Research Fellowship, Imperial College London. P.-J.~Wu acknowledges financial support received through a PhD studentship from the Ministry of Education, Republic of China.

\section*{Acknowledgement}
All authors thank Ester Reina-Torres and Jacques Bertrand for their valuable advice and support provided with the statistical analysis in preparation of this manuscript.   

\section*{Disclosures}
The author have no conflict of interest.


\end{document}